\documentclass[a4paper,12pt]{article}
\usepackage{graphicx}
\begin{document}

\title{{\Large \textbf{Speculative bubbles and fat tail phenomena in a heterogeneous agent model}}} 
\author{Taisei Kaizoji\thanks{E-mail: kaizoji@icu.ac.jp, Home page: 
http://subsite.icu.ac.jp/people/kaizoji/.} \\
\\
Division of Social Sciences, \\
International Christian University\thanks{Division of Social Science, International Christian University, Osawa, Mitaka, Tokyo, 181-8585, Japan.} \\
and \\
Department of Economics, University of Kiel\thanks{Olshausenstr. 40, 24118 Kiel, Germany.}}
\date{}
\maketitle

\begin{abstract}
The aim of this paper is to propose a heterogeneous agent model of stock markets that develop complicated endogenous price fluctuations. We find occurrences of non-stationary chaos, or speculative bubble, are caused by the heterogeneity of traders' strategies. Furthermore, we show that the distributions of returns generated from the heterogeneous agent model have fat tails, a remarkable stylized fact observed in almost all financial markets.\par
\bigskip 
\textbf{Key words: heterogeneous interacting agents, non-stationary chaos, speculative bubbles, leptokurtosis, fat tails}
\end{abstract}

\section{Introduction}

During the past decade, the dynamics of financial asset prices resulting from interaction of heterogeneous agents have become of increasing interest. For example, these price dynamics have been studied Arthur et al. (1997), Brock and Hommes (1997, 1998), Chen and Yeh (2001), Chiarella (1992), Chiarella and He (1999), Chiarella et al. (2001), Day and Huang (1990), De Grauwe et al. (1993), Farmer (2000), Frankel and Froot (1988), Kirman (1991, 1993) and Lux (1995, 1998), Brock, Hommes and Wagener (2001), Gaunersdorfer and Hommes (2000), Kirman and Teyssiere (1998), LeBaron et al. (1999), Lux and Marchesi (1999, 2000) and Youssefmir and Huberman (1997)). However, surprisingly few studies so far have tried to explain prolonged rises in asset prices, known as \textit{speculative bubbles}, due to interaction of heterogeneous agents with different trading strategies and expectations. In contrast, a large number of both theoretical and empirical studies has been made on rational bubbles since Blanchard first proposed this idea in 1979. The aim of this paper is to propose a heterogeneous-agent model that can generate speculative bubbles as a result of the interaction of heterogeneous agents. The theoretical set-up of the model presented here resembles that of previous works in many respects. We think of stock markets as consisting of a {\it market maker} who mediates the trading and generates the market-clearing price, and three typical types of traders that can be observed in financial practice\footnote{These three different classes of traders are often used in the finance literature on heterogeneous-agent systems.}: \textit{fundamentalists} who believe that the asset price will return to the fundamental price, \textit{chartists} who base their trading decisions on an analysis of past price-trends, and \textit{noise traders} who base their trading decisions on noisy information. Traders choose their trading strategy depending on the degree of accuracy of the price prediction. Our model differs from previous works in that the traders' excess demand functions for stock, which are derived from utility maximization, have lower but no upper limits. We demonstrate that this feature is an important factor in speculative bubbles. In particular, we derive the following results: 
\begin{enumerate}
\item When the non-linearity of the excess demand functions is sufficiently strong, a speculative bubble or non-stationary chaos will be observed; that is, irregular fluctuations of the asset price around an upward price trend appear. 

\item When speculative bubbles occur in the stock market, irrational traders, namely, the chartists and noise traders, will drive out of the market rational traders, namely the fundamentalists. 
\item A fat-tailed distribution of the relative stock price change appears. \end{enumerate}

The structure of the paper is as follows: Chapter 2 presents the heterogeneous interacting agent model. Chapter 3 investigates the local and global properties of the deterministic dynamics in the model, and explains why speculative bubbles occur. Chapter 4 shows bubbles of stochastic dynamics and the fat tailed distribution of returns generated from the model. Chapter 5 gives brief concluding remarks.

\section{The Model}

Let us consider a simple stock market with the following characteristics. A large amount of stock is traded. In the market, there are three typical groups of traders with different strategies: fundamentalists, chartists, and noise traders. Traders can invest either in money or in stock. Since the model is designed to describe stock price movements over short periods, such as one day, the dividend from stock and the interest rate for the risk-free asset will be omitted for simplicity. Traders are myopic and bent on maximizing utility. Their utility depends on the price change they expect, and on their \textit{excess demand} for stock rather than simply their demand. Their excess demand is derived from utility maximization. What follows is a more precise account of the decision making of each trader type. 

\subsection{Fundamentalists}

Let $Y_{t}^{f}$ be the amount of money that a fundamentalist holds at time $t $ and let $X_{t}^{f}$ be the number of shares purchased by a fundamentalist at time $t$. Let $p_{t}$ be the price per share of stock at time $t$. The fundamentalist's budget constraint is given by  

\[
Y_{t}^{f}+p_{t}X_{t}^{f}=Y_{t-1}^{f}+p_{t}X_{t-1}^{f},
\]
or equivalently 
\begin{equation}
y_{t}^{f}+p_{t}x_{t}^{f}=0,  
\label{eqn:a1}
\end{equation}

where $y_{t}^{f}=(Y_{t}^{f}-Y_{t-1}^{f})$ denotes the fundamentalist's excess demand for money, and $x_{t}^{f}=(X_{t}^{f}-X_{t-1}^{f})$ his excess demand for stock. Suppose that the fundamentalist's preferences are represented by the utility function,  
\begin{equation}
u=\alpha (y_{t}^{f}+p_{t+1}^{f}x_{t}^{f})+\beta x_{t}^{f}-(1+\beta
x_{t}^{f})\log (1+\beta x_{t}^{f}),\quad \alpha >0,\quad \beta >0,
\label{eqn:a2}
\end{equation}

where $p_{t+1}^{f}$ denotes the fundamentalist's expectation in period $t$ about the price in the following period $t+1$. The parameters $\alpha $ and $ \beta $ are assumed to be positive. Inserting (1) into (2) the fundamentalist's utility maximization problem becomes: 

\begin{equation}
\max_{x^f_t} \quad u = \alpha (p^f_{t+1} - p_t) x^f_{t} + \beta x^f_{t} - (1
+ \beta x^f_{t}) \log (1 + \beta x^f_{t}).  \label{eqn:a3}
\end{equation}

The utility function $u$ satisfies the standard properties: $u^{\prime}(|x^{f}|_{t})>0$, $u^{\prime \prime }(|x^{f}|_{t})<0$ for all $ |x^{f}|_{t}\leq |x^{f\ast }|$ where $|x^{f\ast }|$ denotes the absolute value of $x^{f}$ producing a maximum utility. Thus, the utility function is strictly concave. It depends on the price change expected by fundamentalists $(p_{t+1}^{f}-p_{t})$ as well as the fundamentalist's excess demand for stock $x_{t}^{f}$. The first part $\alpha (p_{t+1}^{f}-p_{t})x_{t}^{f}$ implies that a rise in the expected price change increases his utility. The remaining part expresses his attitude toward risk. Even if the expected price change is positive, he does not want to invest his total wealth in the stock, and vice versa. In this sense, fundamentalists are risk averse. $ \beta $ is the parameter that sets the lower limitation on excess demand. All excess demand for stock derived from the utility maximization is limited to $-1/\beta $. When the expected price change $(p_{t+1}^{f}-p_{t})$ is positive, the maximum value of the utility function is also positive. This means that fundamentalists try to buy stock. By analogy, when the expected price change $(p_{t+1}^{f}-p_{t})$ is negative, the maximum value of the utility function is negative, which means that they try to sell. The utility maximization problem (3) is solved for the fundamentalist's excess demand, 
\begin{equation}
x^f_{t} = \frac{1}{\beta} (\exp(\frac{\alpha (p^f_{t+1} - p_{t})}{\beta}) - 
1).  \label{eqn:a4}
\end{equation}

Excess demand increases as the expected price change $ (p_{t,t+1}^{f}-p_{t})$ increases. It should be noticed that the optimal value of excess supply is limited to $-1/\beta $, while the optimal value of excess demand is not restricted. Since there is little loss of generality in fixing the parameter $\beta $ at unity, below, we will assume $\beta $ to be constant and equal to 1.\par
Then let us think of the fundamentalist's expectation formation. We assume that he form his price expectation according to a simple adaptive scheme:

\begin{equation}
p_{t+1}^{f}=p_{t}+\nu (p^{\ast }-p_{t}).  \label{eqn:a5}
\end{equation}

We see from Equation (5) that fundamentalists believe that the price moves towards the fundamental price $p^{\ast }$ by factor $\nu $. To sum up fundamentalists' behavior: if the price $p_{t}$ is below their expected price, they will try to buy stock, because they consider the stock to be undervalued. On the contrary, if the price is above the expected value, they will try to sell, because they consider the stock to be overvalued. 

\subsection{Chartists}

Chartists are assumed to have the same utility function as the fundamentalists (2). Their behavior is formalized as maximizing the utility function

\begin{equation}
v = \alpha (y_{t} + p^c_{t+1} x^c_{t}) + \beta x^c_{t} - (1 + \beta x^c_{t}) \log (1 + \beta x^c_{t})  
\label{eqn:a6}
\end{equation}

subject to the budget constraint 
\begin{equation}
y^c_{t} + p_{t} x^c_{t} = 0  
\label{eqn:a7}
\end{equation}

where $x_{t}^{c}$ and $y_{t}^{c}$ represent the chartist's excess demand for stock and for money at period $t$, and $p_{t+1}^{c}$ denotes the price expected by him. The chartist's excess demand function for the stock is given by

\begin{equation}
x_{t}^{c}=\frac{1}{\beta }(\exp (\frac{\alpha (p_{t+1}^{e}-p_{t})}{\beta }%
)-1).  \label{eqn:a8}
\end{equation}

His expectation formation is as follows: He is assumed to forecast the future price $p_{t+1}^{c}$ using adaptive expectations, 

\begin{equation}
p^c_{t+1} = p^c_{t} + \mu (p_{t} - p^c_{t}),  \label{eqn:a9}
\end{equation}

where the parameter $\mu $$(0<\mu <1)$ is a so-called error correction coefficient\footnote{The excess demand for chartists $\exp (\beta (p_{t+1}^{e}-p_{t}))-1$ is rewritten as $\exp (\beta (1-mu)(p_{t}^{e}-p_{t}))-1$ under adaptive expectations. Hence, chartists' decisions are based on observation of the past price-data. This type of trader, who simply extrapolates patterns of past prices, is a common stylized example, currently in popular use in heterogeneous agent models (cf. Frankel and Froot (1988) and Brock and Hommes (1997, 1998), and Gaunersdorfer and Hommes (2000)).}. It follows that chartists try to buy stock when they anticipate a rising price for the next period, and, in contrast, try to sell stock when they expect a falling price.

\subsection{Noise traders}

Finally, let us consider the noise traders' decision making. They are assumed to base decisions on noise in the sense of a large number of small events \footnote{For the meaning of noise in financial markets, see Black (1986).}. The behavior of a noise trader can be formalized as maximizing the quadratic utility function

\begin{equation}
W(x^n_{t} , y^n_{t}) = g (y^n_{t} + (p_{t} + \epsilon_t) x^n_{t}) - k
(x^{n}_t)^2  \label{eqn:a10}
\end{equation}

subject to the budget constraint 

\[
y^n_{t} + p_{t} x^n_{t} = 0 
\]

where $x_{t}^{n}$ and $y_{t}^{n}$ represent the noise trader's excess demand for stock and for money at time $t$, respectively. The noise $\epsilon _{t}$ is assumed to be an IID random variable. The excess demand function for stock is given as

\begin{equation}
x^n_{t} = \gamma \epsilon_t, \quad \gamma = \frac{g}{2 k} > 0 
\label{eqn:a11}
\end{equation}

where $\gamma $ denotes the strength of the reaction to noisy information. In short, noise traders try to buy stock if they believe the noise to be good news ($\epsilon _{t}>0$). Inversely, if they believe the noise to be bad news ($\epsilon _{t}<0$), they try to sell it. 

\subsection{The adjustment process of the price}

Let us leave traders' decision-making processes and turn to the adjustment of the stock-market price. We assume the existence of a \textit{market maker}, such as a specialist in the New York stock exchange. The role of the market maker is to give an execution price to incoming orders and to execute transactions\footnote{We don't broach the issue of the market maker's own trading activity because to pursue this would take us beyond the scope of this paper. A heterogeneous agent model that contains a specialist as a market participant has been proposed by Ready (1999). Extension of our model to allow for the market maker's own trading activity is a possible direction for future research.}. 

The market maker announces a price at the beginning of each trading period. Traders then determine their excess demand, based on the announced price and on their expected prices. When the market maker observes either excess demand or excess supply, he applies the so-called {\it short-side rule} to the demands and supplies, taking aggregate transactions for the stock to be equal to the minimum of total supply and demand\footnote{The short side of a market is that where the aggregate volume of desired transactions is smallest: the demand side if there is excess supply, the supply side if there is excess demand.}. Thus traders on the short side of the market will realize their desired transactions. At the beginning of the next trading period, he announces a new price. If the excess demand in period $t$ is positive (negative), the market maker raises (reduces) the price for the following period $t+1$. The process then is repeated. Let $\kappa $ and $\xi $ be the fractions of chartists and of noise traders in the total number of traders, respectively. Then the process of price adjustment can be written as 

\begin{equation}
p_{t+1} - p_{t} = \theta n [(1 - \kappa - \xi) x^f_{t} + \kappa x^c_{t} + 
\xi x^n_{t}],  \label{eqn:a12}
\end{equation}

where $\theta $ denotes the speed of the adjustment of the price, and $ n $ the total number of traders. 

\subsection{Strategy switching}

Interesting questions are whether the rational traders (fundamentalists) will drive irrational traders (chartists and noise traders) out of the market, or whether the irrational traders (chartists and noise traders) will derive the rational traders (fundamentalists) out of the market. As in many studies on heterogeneous interacting agent models (cf. Brock and Hommes (1997, 1998) and Lux (1998)), it may seem natural that switching between different trading strategies plays an important role. In this subsection, we ask the question how a trader makes his choice between the fundamentalist and chartist strategies. The basic idea is that he chooses according to the accuracy of prediction. More precisely, we propose that the proportion of chartists $\kappa _{t}$ is updated according to the difference between the squared prediction errors of each strategy. Formally, we write the dynamics of the proportion of chartists $\kappa $ as 

\begin{equation}
\kappa_t = \frac{(1 - \xi)}{1 + \exp(\psi (E^c_t - E^f_t))},  \label{eqn:a13}
\end{equation}

\[E^c_t = (p_t - p^c_t)^2, \quad E^f_t = (p_t - p^f_t)^2, \]

where $\psi $ measures how sensitively the mass of traders selects the optimal prediction strategy at period $t$. Brock and Hommes (1997, 1998) introduced this parameter $\psi $ as the intensity of choice to switch trading strategies. Equation (13) shows that if the chartists' squared prediction error $E_{t}^{c}$ is smaller than that of fundamentalists $ E_{t}^{f}$, some fraction of fundamentalists will become chartists, and visa versa. 

\section{Deterministic Dynamics}

The aim of this section is to describe the dynamic behavior of the model. In particular, we will indicate local and global properties of the deterministic dynamics when the parameter $\alpha $ is allowed to vary. $ \alpha $ measures the strength of the non-linearity of the excess demand functions. Substituting (4), (5), (8), (9) and (11) to (12), the model can be rewritten as 

\[
p_{t+1}=p_{t}+\theta n[(1-\kappa _{t}-\xi )(\exp (\alpha \nu (p^{\ast}-p_{t}))-1)+\kappa _{t}(\exp (\alpha (1-\mu )(p_{t}^{c}-p_{t}))-1)+\xi \gamma \epsilon _{t}],
\]%
\[
p_{t+1}^{f}=p_{t}+\nu (p^{\ast }-p_{t}),
\]%
\begin{equation}
p_{t+1}^{c}=p_{t}^{c}+\mu (p_{t}-p_{t}^{c}),
\label{eqn:a14}
\end{equation}
where 
\[
\kappa _{t}=\frac{1}{1+\exp (\psi
((p_{t}-p_{t}^{c})^{2}-(p_{t}-p_{t}^{f})^{2}))}.
\]

The first of the equations above represents the adjustment of the price performed by the market maker. The second and third equations show the formation of the expectations of fundamentalists and chartists, respectively. The fourth equation represents the movement of the chartist fraction. Throughout this section, we assume that there exists no noise traders ($\xi =0$). Thus, the dynamics of the system (14) is deterministic. As can be checked easily, the dynamic system (14) has as an unique fixed point: 
$P\equiv (\bar{p},\bar{p}^{f},\bar{p}^{c})=(p^{\ast},p^{\ast },p^{\ast })$.

\subsection{The local stability conditions}

To investigate the dynamics of the model, we shall first determine the local stability region of the unique equilibrium point $P $. The local stability analysis of the equilibrium point $P$ is performed via evaluation of the three eigenvalues of the Jacobian matrix (14) at $P$. Let us denote by

\begin{equation}
c(\lambda) = \lambda^3 - T \lambda^2 + D \lambda 
\label{eqn:a15}
\end{equation}

the associated characteristic polynomial of the Jacobian matrix at $P$, where $ D=(2-\mu -0.5\theta n\alpha (1-\mu +\nu ))$ and $T=(1-\mu -0.5\theta n\alpha (1+\nu )(1-\mu ))$. An eigenvalue of the Jacobian is $0$, and the other roots $\lambda _{1}$ and $\lambda _{2}$, satisfy the relation 

\begin{equation}
\lambda ^{2}-T\lambda +D=0.
\label{eqn:a16}
\end{equation}

Thus, the stability of the equilibrium point is determined by the absolute values of $\lambda _{1}$ and $\lambda _{2}$. The eigenvalues $\lambda _{1}$, $\lambda _{2}$ are $\lambda _{1,2}=T/2+\pm \sqrt{\Delta }/2$ where $\Delta \equiv T^{2}-4D$. As is well known, a sufficient condition for local stability consists of the following inequalities: (i)$1-T+D>0$, (ii) $1+T+D>0 $, and (iii) $D<1$, giving necessary and sufficient conditions for the two eigenvalues to be inside the unit circle of the complex plane. Elementary computations lead to the condition:

\begin{equation}
\alpha < \frac{3 - 2 \mu}{\theta n (1 + \nu - \mu (1 + 0.5 \nu))}.
\label{eqn:a17}
\end{equation}

The above local conditions demonstrate that while increasing $\alpha $ starting from a sufficiently low value inside the stability region, a loss of local stability of the equilibrium point $P$ may occur via a flip bifurcation, when crossing the curve 

\begin{equation}
\alpha = \frac{3 - 2 \mu}{n \theta (1 + \nu - \mu (1 + 0.5 \nu))}.
\label{eqn:a18}
\end{equation}

\subsection{Global bifurcations}

What happens as $\alpha $ is further increased? We will see that first, a stable 2-cycle appears; secondly, the time path of the price diverges; and thirdly, the time evolution of $p_{t}$ displays a remarkable transition from regular to chaotic behavior around an upward time trend. It is important that the non-stationary chaos can be transformed to a stationary series by differentiation once $\Delta p_{t}=p_{t}-p_{t-1}$. The price series fluctuates irregularly around an upward price trend, and the series of the price {\it change} fluctuates chaotically within a finite interval. Below, this non-stationary chaotic pricing will be referred to as a \textit{speculative bubble}. 

Let us look at these phenomena using the numerical simulation. In all numerical simulations below, we will use a set of parameters given by 

\begin{equation}
\theta = 0.001, \quad n = 1000, \quad \mu = \nu = 0.5, \quad \gamma = 1, 
\quad p^* = 100, \quad \psi = 0.001. 
\label{eqn:a19}
\end{equation}

Figure 1 shows a bifurcation diagram with the parameter $ \alpha $ as the bifurcation parameter. The figure summarizes the long-run time evolution of the price change $\Delta p_{t}$, for values of the parameter $3<\alpha <7$. By looking at Figure 1, it is evident that the dynamic behavior starts from a stable point, then goes through a sequence of period doubling bifurcations into a stable cycle of periods $2,4,8,16,...$,and so on.\ Then, the strange attractor set occurs for values beyond the cycle of period three, which appears at about $\alpha =6$. This bifurcation scenario to chaos is the familiar route to chaos, that is to say, the \textit{Feigenbaum scenario}.

\bigskip (Figure 1 goes about here.)
\bigskip

\subsection{Speculative bubbles}

Why does a speculative bubble or non-stationary chaos appear when $\alpha $ increases? Unfortunately, it is difficult to prove mathematically the existence of the strange attractor. However, as we shall see, it is possible to give an intuitive interpretation of speculative bubbles. To begin with, let us consider why chaotic behavior of the price change is observed for large values of $\alpha $. Figure 2 indicates a phase plot ($\Delta p_{t},\Delta p_{t+1}$) of the sequence of price change $\Delta p_{t}$ for $ \alpha =5.5$. The phase plot is constructed by iteration of 20000 times, but here we use just the last 10000 iterations. A glance at Figure 2 shows a fixed point on the phase curve, depicted as an intersection between the phase curve and the 45-degree line\footnote{We can also show that the phase curve is stretched upwards when $\alpha $ increases.}. As can be traced in this phase plot, a set of states of $\Delta p_{t}$ initially located close to the fixed point will first be stretched, and then at some point will folded back on itself. This stretching and folding is the root of the chaotic price behavior (For mathematical proof see Li and York (1975), and Marotto (1978).).

\bigskip (Figure 2 goes about here.)

\bigskip

The next question is why speculative bubbles occur when $\alpha $ is large. The price adjustment equation suggests that a price trend can occur only when excess demand aggregated over two types of traders is positive on average. Why would average excess demand be positive as $\alpha $ become large? For an answer, let us recall the traders' excess-demand functions (4) and (8). $x_{t}$ denotes the traders' excess demand, and $p_{t+1}^{e}$ denotes the expected price. The partial derivative of the excess-demand function with respect to the parameter $\alpha $ is obtained as

\begin{equation}
\frac{\partial x_{t}}{\partial \alpha }=(p_{t+1}^{e}-p_{t})\exp (\alpha (p_{t+1}^{e}-p_{t})),
\label{eqn:a20}
\end{equation}
where we assume $\beta =1$. This equation suggests a rise in $\alpha $ increasing excess demand exponentially, given the expected price change $(p_{t+1}^{e}-p_{t})$ is positive. As shown in the preceding subsection, the maximum value of the excess demand increases freely when $\alpha $ increases, while the maximum value of the excess supply never can be lower than $-1/\beta $, regardless of the value of $\alpha $. Assume that, in the initial state, the price $ p_{t}$ exceeds the fundamental price $p^{\ast }$ in period $t$, and that the chartists forecast a rise over the next period. The fundamentalists predict a falling price. Then, chartists try to buy and fundamentalists try to sell stock. When $\alpha $ is sufficiently large, one can safely state that chartist excess demand will exceed fundamentalist excess supply. The market maker observes excess demand, and raises the price for the following period. If the rise is strong, chartists may predict a new rise, which provokes yet another reaction of the market maker. Once the price deviates strongly from the fundamental price, the fundamentalists become perpetual sellers of stock, and when $\alpha $ is sufficiently large, the chartists' excess demand may begin to exceed the fundamentalists' excess supply on average. Furthermore, fundamentalists may be driven out of the market by evolutionary competition expressed by (13)\footnote{In the following section, we shall be examining this subject.}. It follows that when the value of $\alpha $ is sufficiently large, average excess demand is positive and speculative bubbles occur. 

\section{Stochastic dynamics}

To adjust the model to approximate real stock markets, we now assume the existence of noise traders. Below, we set $\xi =0.3,$ i.e., 30 percent of the total number of traders are noise traders. Figure 3a and Figure 3b indicate the time series of the price and of the price change for $\alpha =5.5$, disturbed by dynamic noise with $\epsilon _{t}\sim N(0,1)$. The figures show that the time series of the price has an upward trend and that it fluctuates irregularly\footnote{Many numerical simulations for various values of $\alpha $ and $\xi $ suggest that the existence of noise traders does not change essentially the properties of the price dynamics, such as speculative bubbles.}. 
\bigskip (Figure 3a and Figure 3b go about here.)

\subsection{Evolutionary competition}

Another very interesting question is whether the irrational traders (chartists and noise traders) can survive competition against the rational traders (fundamentalists), or whether the rational traders (fundamentalists) can survive competition against the irrational traders (chartists and noise traders). Figure 4a illustrates the time series of the fraction of chartists $\kappa $ for the parameters $\alpha =4$ and $\xi =0.3$. The speculative bubble does not occur under this set of parameters. Figure 9a shows that $\kappa _{t}$ continues to fluctuate around 0.35. This means that, on average, the same number of fundamentalists and chartists exists, and that the irrational traders will stay in the market forever. On the other hand, Figure 4b shows the time series of the fraction of chartists $\kappa _{t}$ for the parameters $\alpha =5$ and $\xi =0.3$. There occurs a speculative bubble under this set of parameters. We see from Figure 9b that $ \kappa _{t}$ finally converges to $0.7$. This means that fundamentalists die out, and irrational traders survive competition. The reason is the speculative bubble caused by the strong non-linearity of the excess-demand functions. When speculative bubbles occur, the price continues to deviate from the fundamental price, so that the prediction errors of the fundamentalists' trading strategy increase compared to those of the chartists' trading strategy. Therefore, the number of fundamentalists decreases and the number of chartists increases. 

\bigskip (Figure 4a and Figure 4b go about here.)

\subsection{Fat tail phenomenon}

A number of empirical studies have found that the empirical distributions of returns in almost all financial data show an uni-modal bell shape, and possess excessive fourth moments, so-called \textit{leptokurtosis}, (cf. Pagan (1996)). Figure 5 is a histogram of the stock-return series \footnote{Returns are defined as $r_{t}=\log p_{t+1}-\log p_{t}$.} for $\alpha =4$. It is constructed by 20000 iterations of the system (14) with $\alpha =5.5$. The shape of the histogram appears well behaved: an uni-modal bell-shape. Table 1 presents sample statistics of the stock returns, which are based upon several series of 20000 data points. The sixth and seventh columns denote the dynamic properties of the price and the price change, respectively. The mean value is very close to zero, and the standard deviation is small. The skewness is positive, and the kurtosis is significantly positive. The important result is that the histogram exhibits apparently more probability mass in the tails and the centre than the standard Normal. Therefore, the time series of the returns are leptokurtotic, i.e., possessing excessive fourth moments. \newline

(Figure 5 goes about here.)\newline

\begin{table}[htbp]
\caption[Statistics]{Statistics of returns from artificial data}
\begin{center}
\begin{tabular}{cccccccc}
\hline
$\alpha $ & \textit{Mean} & \textit{S.D.} & \textit{Skewness} & \textit{%
Kurtosis} & \textit{Price} & \textit{Return} & $\rho $ \\ \hline
1 & 0 & 0.014 & 0.023 & 0.061 & stable & stable & 4.83 \\ 
2 & 0 & 0.016 & 0.210 & 0.323 & stable & stable & 4.14 \\ \hline
3 & 0 & 0.02 & 0.703 & 1.552 & periodic & periodic & 3.2 \\ 
4 & 0 & 0.026 & 1.411 & 4.277 & periodic & periodic & 2.32 \\ \hline
5 & 0 & 0.004 & 2.963 & 17.605 & bubble & chaotic & 4.58 \\ \hline
\end{tabular}%
\end{center}
\end{table}

However, the kurtosis is a relatively limited measure of deviation for Gaussian statistics. Recent literature finds that the distribution of returns is not only leptokurtotic, but also belongs to the class of \textit{fat-tailed} distributions. This finding is generally accepted as a universal characteristic of practically all 
financial returns. More formally, it has been shown that the tails of the distribution of returns follow approximately a power law:

\begin{equation}
F(|z_t| > x) \sim c x^{- \rho} 
\label{eqn:a21}
\end{equation}

where the function $F$ denotes the cumulative probability distribution of the normalized stock returns and $\rho $ is called tail index. We normalized the stock returns, 
\begin{equation}
z_t = \frac{(r_{t} - \langle r_t \rangle)}{\upsilon} 
\label{eqn:a22}
\end{equation}

where $\langle r_{t}\rangle $ denotes the time average and $\upsilon $ denotes the standard deviation. The last column of Table 1 shows the values of the tail index $\rho $ for tail sizes up to 5 percent of the size of the return series generated from our model under various values of $ \alpha $\footnote{The values of $\rho $ are computed by the Hill tail index estimate, which is a standard working tool for estimations of the Pareto exponent of the tails.}. Most empirical studies show estimates of $\rho $ falling in the range of 2 to 4 (cf. Pagan (1996), Gopikrishnan et al. (1998)). In Table 1 the values of $ \rho $ range between 2 and 4, as long as periodic solutions of the price are observed. Thus, we find a power-law behavior for the cumulative distribution of the stock returns with $\rho $ well outside the Levy regime ($0<\rho <2$). Consequently, the time series of the returns generated from the above artificial stock market share this remarkable feature of actual data from stock markets. 

\section{Concluding Remarks}

In this paper, we proposed a heterogeneous interacting agent model for stock markets. It generates three important findings: (1) the non-linearity of the excess demand functions derived from the traders' optimization behavior might generate speculative bubbles; (2) fundamentalists may be driven out of the market by irrational traders, such as chartists and noise traders, provided a speculative bubble occurs; and (3) the return distributions derived from the model have fat tails, an important characteristic of empirical financial data which is observed in almost all financial markets. There are, however, remarkable differences between the time series generated by our model and actual financial data. For example, the noisy time series of the price change (Figure 3b) has strongly significant autocorrelations at the first two lags ($0.5$ and $0.1$), whereas in reality the series of price changes or returns have almost no significant autocorrelations at any lags. Furthermore, the returns series generated by the model exhibit no volatility clustering, which is a universal property of real returns series. Apparently, our model is too simple to explain all stylized facts of real financial data at the same time and, thus, needs to be further developed. This we leave to future work. 

\section{Acknowledgements}

I would like to thank Thomas Lux for helpful suggestions and encouragement. Comments by Tatsuhiko Aoki, Shu-Heng Chen, Janusz Holyst, Akio Matsumoto, Tetsuya Misawa, Tamotsu Onozaki, Peter Flaschel, Reiner Frank, Willi Zemmler, Toichiro Asada, and anonymous referees were very helpful and have led to several improvements. Financial support by the Alexander Humboldt Foundation, and the Japan Society for Promotion of Science under the Grant-in-Aid, No. 06632 gratefully acknowledged.  All remaining errors, of course, are mine.

\newpage


\begin{figure}
\begin{center}
  \includegraphics[height=19cm,width=14cm]{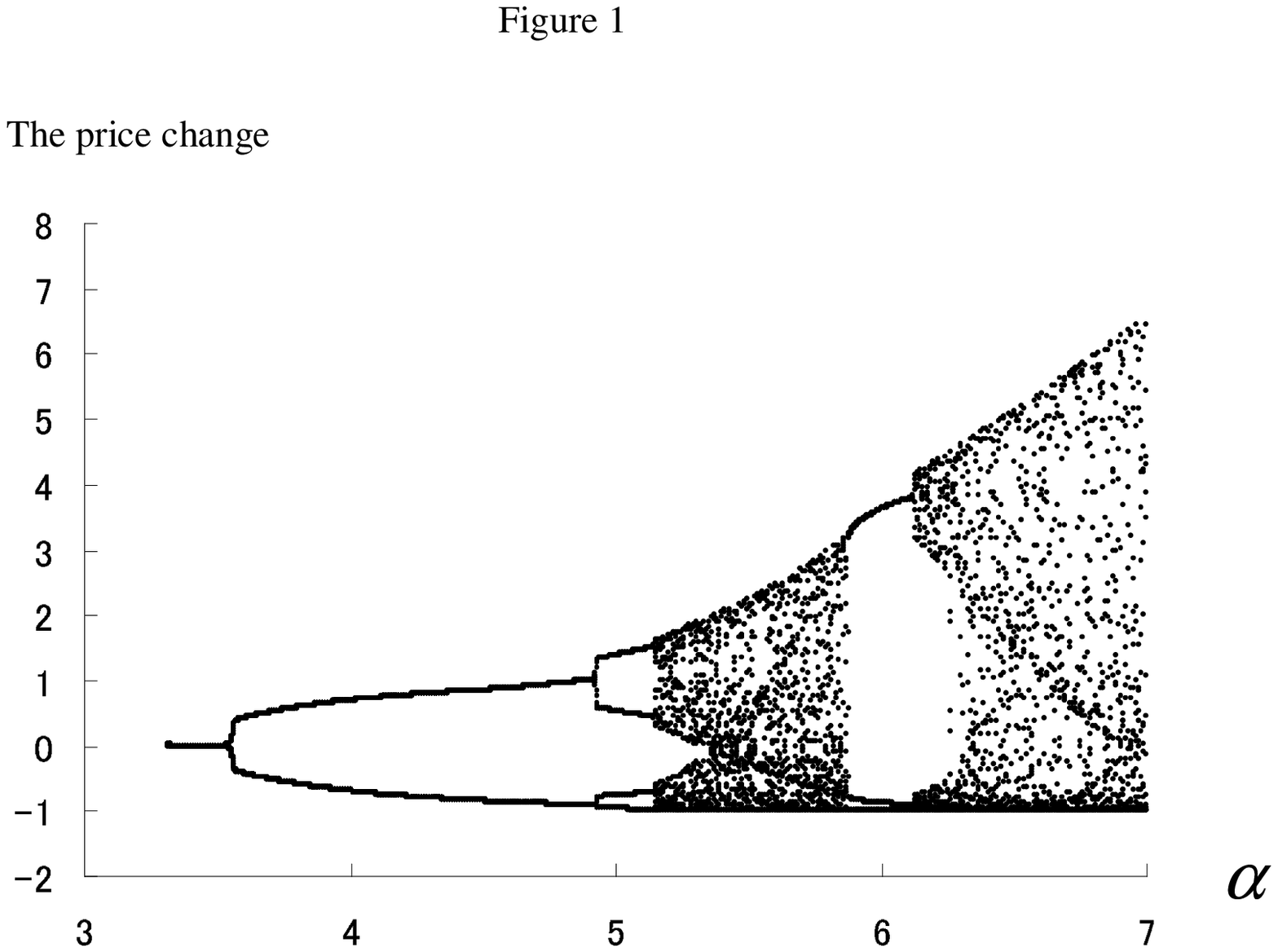}
\end{center}
\caption{The bifurcation diagram of the stock return when the parameter $ \alpha $ varies smoothly from $3 $ to $7 $. The other parameter values: $ \theta = 0.001 $, $n = 1000 $, $\mu = \nu = 0.5 $, $\gamma = 1 $, $p^* = 100 $, $\psi = 0.001 $, and $\xi = 0 $.}
\label{fig1}
\end{figure}

\begin{figure}
\begin{center}
  \includegraphics[height=19cm,width=14cm]{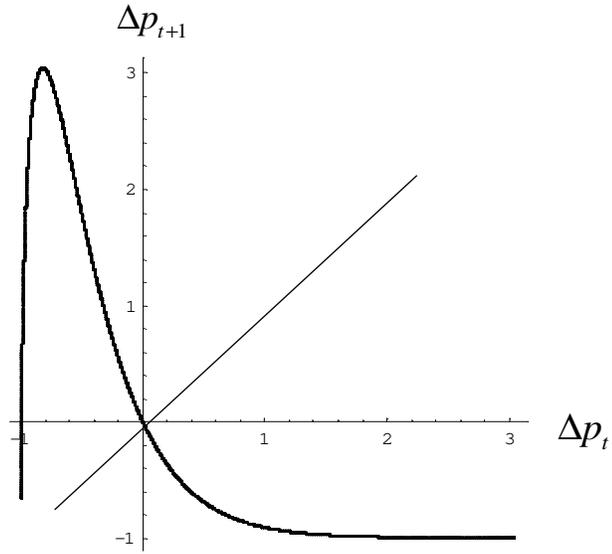}
\end{center}
\caption{ A phase plot ($\Delta p_t, \Delta p_{t-1} $) of the sequence of price change $\Delta p_t $ generated for $\alpha = 5.5 $. The other parameter values are: $\theta = 0.001 $, $n = 1000 $, $\mu = \nu =0.5 $, $\gamma = 1 $, $p^* = 100 $, $\psi = 0.001 $, and $\xi = 0 $.}
\label{fig2}
\end{figure}

\begin{figure}
\begin{center}
  \includegraphics[height=19cm,width=14cm]{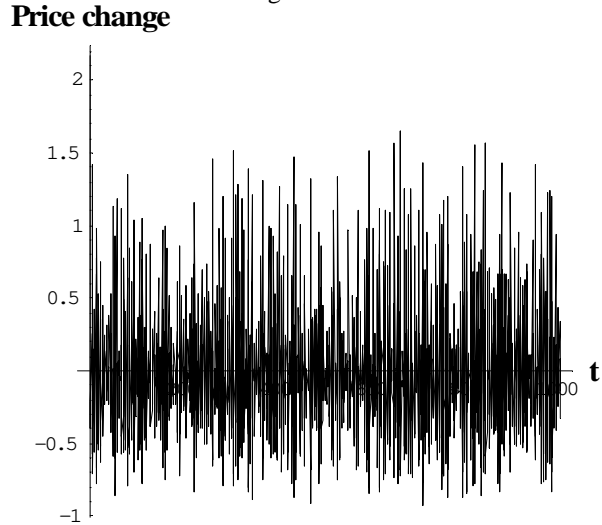}
\end{center}
\caption{(a) The time series of the price for $\alpha = 5.5 $, buffeted with dynamic noise with $\epsilon \sim N(0,1) $. The other parameter values are: $\theta = 0.001 $, $n = 1000 $, $\mu = \nu = 0.5 $, $ \gamma = 1 $, $p^* = 100 $, $\psi = 0.001 $, and $\xi = 0.3 $. (b) The time series of the price change for $\alpha = 5.5 $, buffeted with dynamic noise with $\epsilon \sim N(0,1) $. The other parameter values are: $\theta = 0.001 $, $n = 1000 $, $\mu = \nu = 0.5 $, $ \gamma = 1 $, $p^* = 100 $, $\psi = 0.001 $, and $\xi = 0.3 $. } 
\label{fig3}
\end{figure}

\begin{figure}
\begin{center}
  \includegraphics[height=19cm,width=14cm]{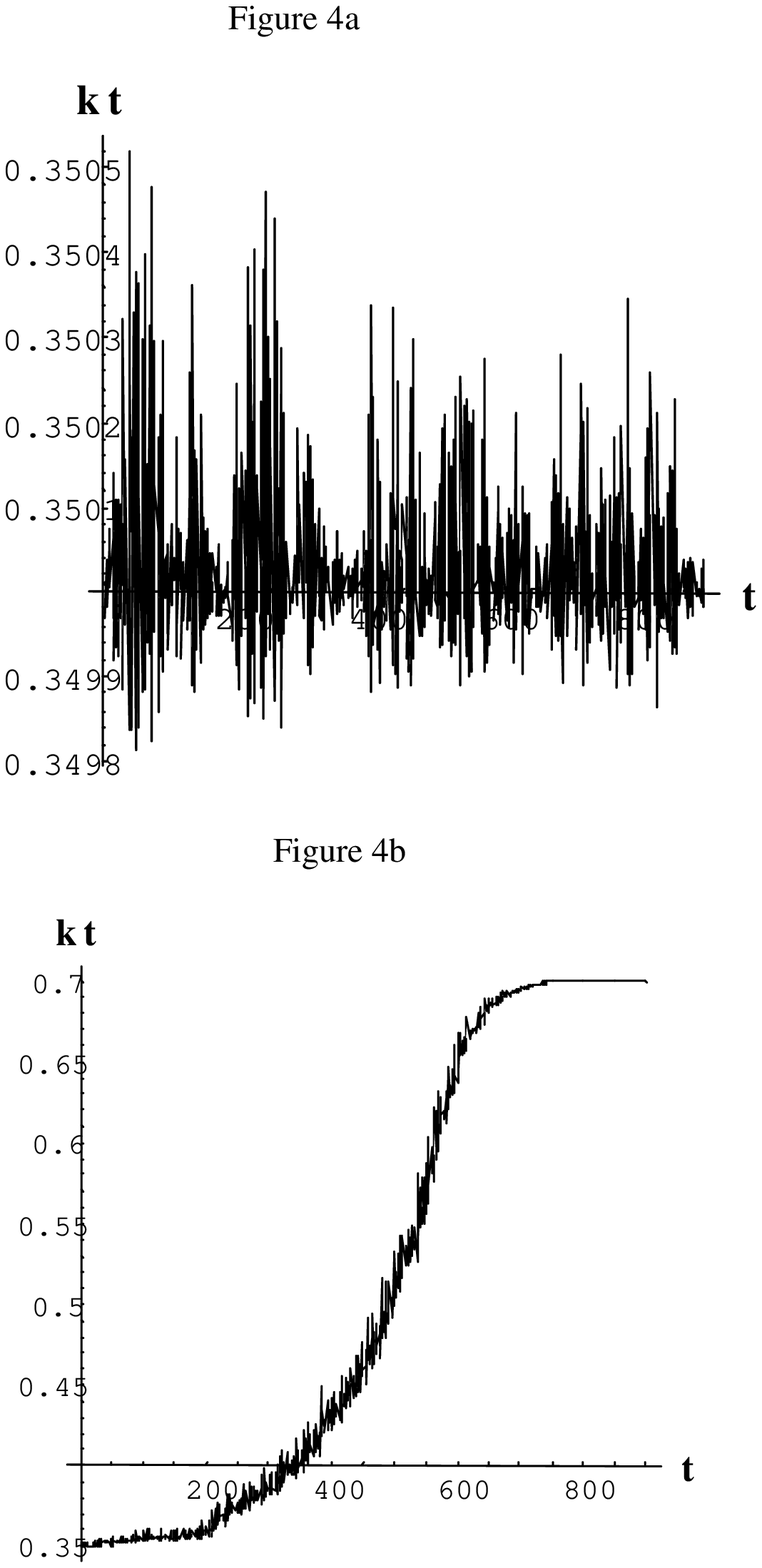}
\end{center}
\caption{(a) The time-series plots of the fraction of the chartists $\kappa $ for the parameters $\alpha = 4 $ and $\xi = 0.2 $. The other parameter values are: $\theta = 0.001 $, $n = 1000 $, $\mu = \nu = 0.5 $, $ \gamma = 1 $, $p^* = 100 $, $\psi = 0.001 $, and $\xi = 0.3 $. (b) The time series plots of the fraction of the chartists $\kappa $ for the parameters $\alpha = 5 $ and $\xi = 0.3 $. The other parameter values: $\theta = 0.001 $, $n = 1000 $, $\mu = \nu = 0.5 $, $ \gamma = 1 $, $p^* = 100 $, $\psi = 0.001 $, and $\xi = 0.3 $. } 
\label{fig4}
\end{figure}

\begin{figure}
\begin{center}
  \includegraphics[height=19cm,width=14cm]{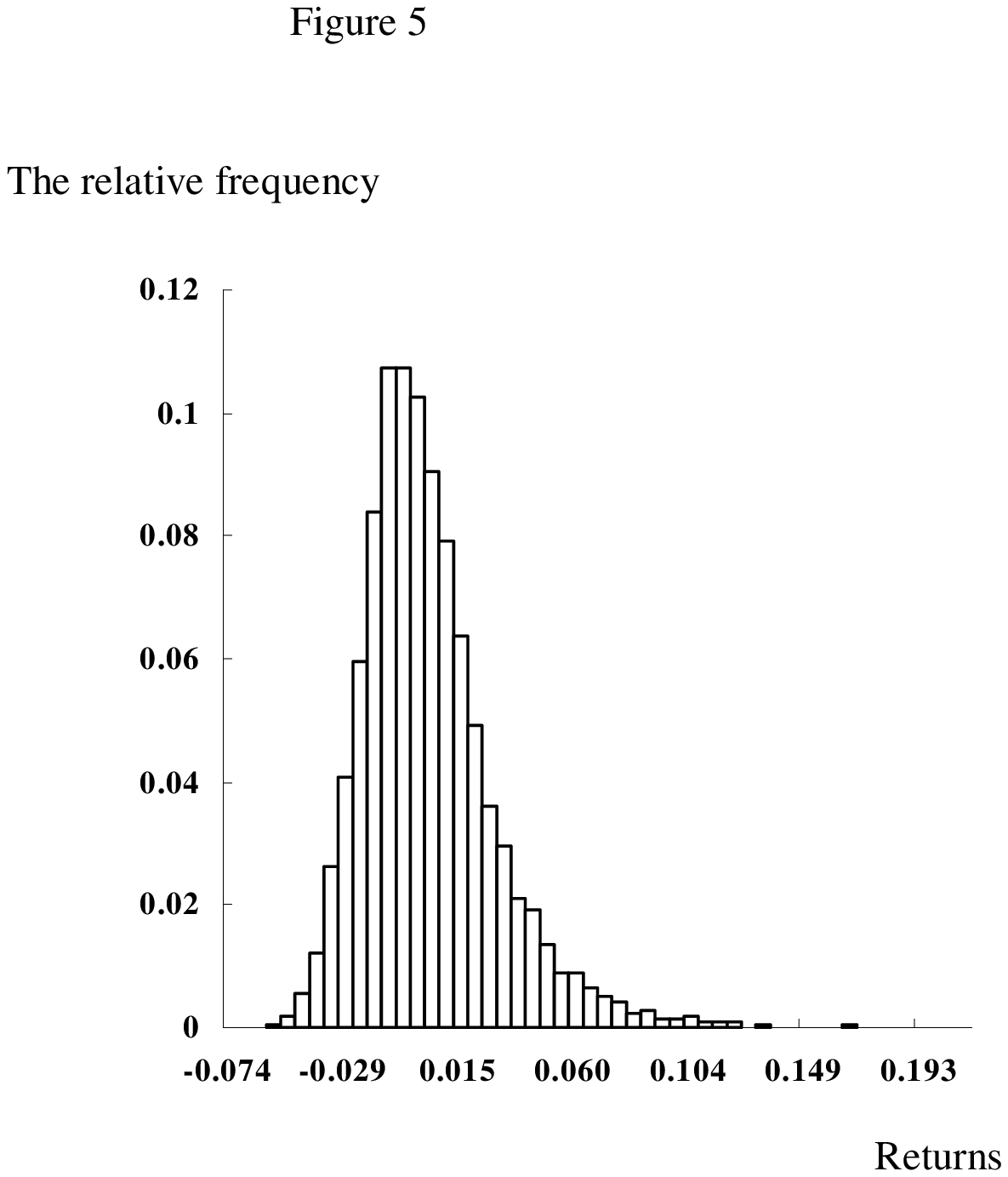}
\end{center}
\caption{ A histogram of returns generated from the time series
of the price for $\alpha = 5.5 $. The other parameter values are: $\theta =
0.001 $, $n = 1000 $, $\mu = \nu = 0.5 $, $\gamma = 1 $, $p^* = 100 $, $\psi
= 0.001 $, and $\xi = 0.3 $.} 
\label{fig5}
\end{figure}
\end{document}